\documentclass[a4paper,superscriptaddress,aps,prb,twocolumn,floatfix,citeautoscript]{revtex4}

\makeatletter
\def\@dotsep{4.5}
\makeatother

\usepackage{amsmath,amsbsy,amsfonts,mathrsfs,graphicx,sidecap,bm}

\newcommand{\I}{\mathrm{i}}
\newcommand{\D}{\mathrm{d}}

\renewcommand{\vec}[1]{{\bf #1}}

\newcommand{\mint}[1]{\int\! \D^{3} #1 \, }

\newcommand{\ba}{\begin{array}}
\newcommand{\ea}{\end{array}}

\newcommand {\lsi}{Laboratoire des Solides Irradi\'es,
  CNRS-CEA-\'Ecole Polytechnique, Palaiseau, France}

\newcommand {\coimbra}{Centro de F\'{\i}sica Computacional, Departamento de F{\'\i}sica, Universidade de
  Coimbra, Coimbra, Portugal}

\newcommand{\berlin}{Institut f{\"{u}}r Theoretische Physik, Fachbereich Physik der
Freie Universit{\"{a}}t Berlin, Arnimallee 14, D-14195 Berlin, Germany}

\newcommand{\cagliari}{INAF Osservatorio Astronomico di Cagliari, Astrochemistry Group,
Strada n.54, Loc. Poggio dei Pini, I09012 Capoterra (CA), Italy}

\newcommand {\etsf}{European Theoretical Spectroscopy Facility}

\begin{document}

\title{
Efficient calculation of van der Waals dispersion coefficients with time-dependent
density functional theory in real time: application to
polycyclic aromatic hydrocarbons
}

\date{\today}

\author{Miguel A. L. Marques}
\email{marques@teor.fis.uc.pt}
\affiliation{\coimbra}
\affiliation{\etsf}

\author{Alberto Castro}
\affiliation{\berlin}
\affiliation{\etsf}

\author{Giuliano Malloci}
\affiliation{\cagliari}

\author{Giacomo Mulas}
\affiliation{\cagliari}

\author{Silvana Botti}
\affiliation{\lsi}
\affiliation{\coimbra}
\affiliation{\etsf}

\begin{abstract}
  The van der Waals dispersion coefficients of a set of polycyclic
  aromatic hydrocarbons, ranging in size from the single-cycle benzene
  to circumovalene (C$_{66}$H$_{20}$), are calculated with a real-time
  propagation approach to time-dependent density functional theory
  (TDDFT). In the non-retarded regime, the Casimir-Polder integral is
  employed to obtain $C_6$, once the dynamic polarizabilities have been
  computed at imaginary frequencies with TDDFT. On the other hand, the
  numerical coefficient that characterizes the fully retarded regime
  is obtained from the static polarizabilities. This {\em ab initio}
  strategy has favorable scaling with the size of the system -- as
  demonstrated by the size of the reported molecules -- and can be
  easily extended to obtain higher order van der Waals coefficients.
\end{abstract}

\maketitle

\section{INTRODUCTION}

When two molecules are far apart, they still interact through long
range electromagnetic forces, named after J. D. van der
Waals.\cite{vanderWaals-1869} These are crucial to understand a wide
range of phenomena: the chemistry of rare gases,\cite{christe-2001}
protein folding and dynamics,\cite{roth-1996} the machinery of
neurotransmitters\cite{siegel-2006} and the molecular chemistry in the
interstellar medium\cite{naray-szabo-1990} are just some examples.
In particular, in astrochemistry van der Waals forces are essential 
to describe neutral-neutral reactions, which are not yet well accounted
for in rate-reaction models.
These intermolecular forces are mainly originated by electric
multipole-multipole interactions. Even if the static multipoles of the
molecules are null (as the dipole of a single spherical atom in the
ground state), they still contribute to the interaction due to their
spontaneous oscillations.  These contributions of dynamical origin,
that cannot be explained by Electrostatics, are the London dispersion
forces.\cite{london-1936} These are attractive interactions, and arise
from the mutual polarization of the two electronic clouds. They
dominate the long range dynamics of the molecules, and are also
present at short distances -- although masked by other stronger (ionic
or covalent) forces.

The calculation of these dispersion forces can be a challenging
problem.\cite{dobson-2001} Three regimes have to be distinguished,
according to the distance that separates the interacting molecules:

(i)~{\it short distances}, such that there is non-negligible overlap
of the electronic clouds of the two molecules. This is the most
difficult situation, since it requires, in principle, a {\em
supermolecule} calculation, i.e. the treatment of the two molecules
combined together as a single entity. The valid approaches in this
overlapping regime (and therefore also valid in the non-overlapping
regime) are sometimes called seamless van der Waals
techniques. Numerous possibilities exist, although none of them
entirely satisfactory: full configuration interaction for very small
molecules,\cite{lee-2001} M{\o}ller-Plesset perturbation
theory\cite{rowley-2006} or Monte Carlo methods.\cite{anderson-1993}
Attempts to use ground state density functional theory (DFT) are
specially challenging;\cite{dobson-2001} In the realm of
time-dependent density-functional theory (TDDFT), the recent work of
Dobson~\cite{dobson-2006} summarizes the current research, largely
based on the adiabatic connection / fluctuation dissipation theorem.
Dispersive forces can also be added to standard DFT through empirical correction
terms. These, however, require the previous knowledge of the $C_6$ van der Waals coefficients 
(see below), often roughly estimated from the atomic $C_6$'s .\cite{grimme-2004-2005}
In the present work we will not consider this regime.

(ii)~{\it long distances}, such that we can neglect the overlap.  In
this case, the electrons belonging to different molecules are
distinguishable, and one can isolate the Coulomb operator that
corresponds to interactions between electrons of different molecules,
and apply (second order) perturbation theory for this
operator.\cite{jeziorsky-1994} The first term in the perturbative
expansion of the interaction energy decays as $-C_6/R^6$, where $R$ is
the intermolecular distance.

(iii)~{\it very long distances}, such that retardation effects become
important.\cite{casimir-1948} This means that the time it takes for
the photons that mediate the electromagnetic interaction to travel
between the two molecules is not negligible. Retardation is described
by a correction factor that is equal to unit for small distances and
proportional to $1/R$ for large distances.

The knowledge of the various static and dynamic multipole
polarizabilities at imaginary frequencies suffices to compute the van
der Waals interaction in the regimes (ii) and (iii). These response
functions can be calculated within a variety of quantum chemical
methods, typically through the evaluation of a wide range of
excitation energies and associated transition matrix
elements. Alternatively, at a lower computational cost, we can use
time-dependent density-functional theory
(TDDFT).\cite{runge-1984,tddft-book} This theory is very successful in
predicting optical spectra (i.e., essentially the imaginary part of
the dynamic polarizability at real frequencies) for finite systems,
and has been used to study a wealth of molecules and
clusters.\cite{castro-2004} The first calculation of $C_6$
coefficients using TDDFT was performed in 1995 by van Gisbergen et al
for a variety of small molecules.\cite{vanGisbergen-1995}

TDDFT in the perturbative regime can be formulated in a variety of
flavors.\cite{marques-2006} With the purpose of calculating $C_6$
coefficients, one can find several approaches. Van Gisbergen et
al\cite{vanGisbergen-1995} opted for a self-consistent calculation of
the response function. The matrix formulation of linear response
theory\cite{casida-1995} can also be employed, yielding oscillator
strengths and excitation energies, sufficient ingredients for the
computation of the dynamic polarizability. A slightly different
approach is the polarization propagator
technique,\cite{oddershede-1984} which can be constructed on top of
TDDFT (or on top of other electronic structure theories,
e.g. time-dependent Hartree-Fock, TDHF). The linear polarization
propagator was demonstrated to work for complex
frequencies\cite{norman-2001}, opening the possibility of computing
$C_6$ coefficients.\cite{norman-2003} With this technique, Jiemchooroj
and collaborators computed $C_6$ coefficients of noble gases
atoms,\cite{norman-2003} $n$-alkanes\cite{norman-2003} ($n\le 7$),
polyacenes,\cite{jiemchooroj-2005} the C$_{60}$
molecule,\cite{jiemchooroj-2005} and sodium clusters up to 20
atoms\cite{jiemchooroj-2006}.  For large molecules, Banerjee and
Harbola have proposed the use of orbital free
TDDFT,\cite{banerjee-2002} providing satisfactory results for large
sodium clusters.

In this work, we propose an alternative scheme, based on the explicit
propagation of the time-dependent Kohn-Sham
equations.\cite{yabana-1996,yabana-1999} This approach has already
proved its usefulness to calculate the dynamic polarizability of large
systems (see, e.g., Ref.~\onlinecite{lopez-2005}). The explicit
time-dependent picture, where one is confronted with an initial value
problem, can be preferable over the the time-independent picture --
where the mathematical problem is diagonalization --, especially
regarding scalability with the size of the system.  For a discussion
in the field of TDDFT, see Ref.~\onlinecite{marques-2006}.

We have chosen an extensive set of polycyclic aromatic hydrocarbons
(PAHs) in order to validate the methodology.  PAHs, a large class of
conjugated $\pi$-electron systems, are of great importance in many
areas, among which combustion and environmental chemistry, materials
science, and astrochemistry. PAHs are found in carbonaceous meteorites, 
in interplanetary dust particles, and are thought to be the most abundant
molecular species in space after H2 and CO, playing a crucial role in
the energy and ionization balance of interstellar matter in galaxies.
Motivated by this astrochemical relevance, some of us have recently performed an
extensive study of these systems.\cite{malloci-2004,malloci-2007-II}
This research is collected in a thorough compendium of molecular
properties of PAHs.\cite{malloci-2007-II} The computation of van der
Waals constants for pairs of PAHs, as a first approach to the analysis
of their intermolecular properties, is therefore a natural extension
of this work. In fact, van der Waals parameters are a crucial ingredient to model condensation 
and evaporation of PAH clusters, which are another important player in the 
physics/chemistry of the interstellar medium. In current works,
people usually employ empirical van der Waals parameters for the relatively long-range part 
of the interaction in molecular dynamics simulations,
together with some tight-binding approximation for the short-range part.
\cite{rapacioli-2005-2006}

One previous calculation of dispersion coefficients of PAHs, based on
the complex linear polarization propagator method, was reported in
Ref.~\onlinecite{jiemchooroj-2005}, although limited to benzene,
naphthalene, anthracene, and naphtacene. This fact adds a further
motivation to our study, since it permits to compare the results of
the two schemes.

\section{METHODOLOGY}

When both the wavefunction overlap and the retardation effects are
almost null [situation (ii)], the application of second order
perturbation theory leads to an expansion of the interaction energy
with respect to the inverse of the intermolecular distance $(1/R)$:
\begin{equation}
  \label{eq:eofr}
  \Delta E(R) = -\sum_{n=6}^{\infty} \frac{C_n}{R^{n}}\,.
\end{equation}
The coefficients $C_n$ are usually called Hamaker
constants.\cite{hamaker-1937} The leading non-null term $C_6$ is due
to the dynamic dipole-dipole polarizability. The odd terms until
$n=11$ are also null for spherical molecules (or if an average over
relative orientations is taken), if we neglect retardation effects. In
principle, the coefficients also depend on the relative orientation of
the molecules.

The $C^{AB}_6$ dispersion Hamaker constant for a pair of molecules $A$
and $B$, averaged over all possible relative orientations, is related
to the dipole molecular polarizability through the Casimir-Polder
relation (atomic units will be used hereafter):
\begin{equation}
  \label{eq:c6}
  C^{AB}_6 = \frac{3}{\pi}\int_0^{\infty} 
    \!\!\!{\rm d}u \;\alpha^{(A)}({\rm i}u)\;\alpha^{(B)}({\rm i}u)\,,
\end{equation}
where $\alpha^{(X)}({\rm i}u)$ is the average of the dipole
polarizability tensor of molecule $X$,
$\boldsymbol{\alpha}^{(X)}$, evaluated at the complex frequency ${\rm
i}u$:
\begin{equation}
  \alpha^{(X)}({\rm i}u) = \frac{1}{3} {\rm Tr} [\boldsymbol{\alpha}^{(X)}({\rm i}u)]\,.
\end{equation}
It should be noted that: (i) If we fix the relative orientation of the
molecules, the orientation dependent Hamaker constant can also be
calculated from the $\boldsymbol{\alpha}$ tensors, considering the
appropriate linear combination of their components; (ii) higher order
Hamaker constants, useful for shorter distances, can be obtained
through the use of analogous formulae involving higher order multipole
polarizability tensors. The expressions accounting for these two
generalizations can be found, for example, in
Ref.~\onlinecite{jeziorsky-1994}. The calculations presented below are
solely concerned with Eq.~\eqref{eq:c6}; however we wish to stress
that the methodology trivially yields full polarizability tensors of
arbitrary order, hence providing the possibility to tackle those general
cases, with almost no extra computational cost.

On the other hand, when the distance increases and retardation effects
become important, the van der Waals interaction depends solely on the
static polarizability,\cite{casimir-1948,calbi-2003} decaying as
$\Delta E(R) = -K^{AB}/R^7$, where the constant is
\begin{equation}
  \label{eq:kab}
  K^{AB} = \frac{23 c}{8 \pi^2} \alpha^{(A)}(0)\; \alpha^{(B)}(0)
  \,,
\end{equation}
where $c$ is the velocity of light in vacuum.

The TDDFT time propagation approach to obtain the dynamic
polarizability components can be summarized as follows: let us assume
a weak electrical (spin-independent) dipole perturbation
\begin{equation}
  \delta v_{{\rm ext},\sigma}(\vec{r},\omega) =  - x_j \kappa(\omega)\,.
\end{equation}
This defines an electrical perturbation polarized in the direction $j$: $\delta
\vec{E}(\omega) = \kappa(\omega)\hat{e}_j$.  The response of the system dipole
moment in the $i$ direction
\begin{equation}
  \label{eq:dipole_response}
  \delta \langle \hat{X}_i\rangle(\omega) =
  \sum_{\sigma}\mint{r} x_i \; \delta n_\sigma(\vec{r},\omega)
\end{equation}
is then given by:
\begin{equation}
  \delta \langle \hat{X}_i\rangle(\omega) = -\kappa(\omega) \sum_{\sigma\sigma'}
  \mint{r}\!\!\mint{r'} x_i\;\chi_{\sigma\sigma'}(\vec{r},\vec{r}',\omega)\; x'_j\,,
\end{equation}
where $\chi_{\sigma\sigma'}$ is the response function of the system:
\begin{equation}
\delta n_\sigma(\vec{r},\omega) = 
\sum_{\sigma\sigma'}\mint{r} \chi_{\sigma\sigma'}(\vec{r},\vec{r}',\omega) 
\delta v_{{\rm ext},\sigma'}(\vec{r}',\omega)\,.
\end{equation}
The dynamic dipole polarizability
$\alpha_{ij}(\omega)$ is the quotient of the induced dipole moment in
the direction $i$ with the applied external electrical field in the
direction $j$, which yields:
\begin{equation}
  \label{eq:polarizability_def}
  \alpha_{ij}(\omega) =
  -\sum_{\sigma\sigma'}\mint{r}\!\!\mint{r'} x_i\;
  \chi_{\sigma\sigma'}(\vec{r},\vec{r}',\omega)\; x'_j\,.
\end{equation}
The dynamic polarizability elements may then be arranged to form a
second-rank symmetric tensor, $\boldsymbol{\alpha}(\omega)$.  We
consider now a sudden external perturbation at $t=0$ (delta function
in time, which means $\kappa(\omega) = \kappa$, equal for all
frequencies), applied along a given polarization direction, say
$\hat{e}_j$.  By propagating the time-dependent Kohn-Sham
equations,\cite{castro-2004-II} we obtain $\delta n(\vec{r}, t)$.
The polarizability element $\alpha_{ij}(\omega)$ may then be calculated
easily via:
\begin{eqnarray}
\nonumber
\alpha_{ij}(\omega) & = & -\frac{\delta \langle \hat{X}_i\rangle(\omega)}{\kappa} =
-\frac{1}{\kappa}\mint{r} x_i\; \delta n(\vec{r}, \omega) = 
\\
& & -\frac{1}{\kappa} \int\!\!{\rm d}t \mint{r} x_i\; \delta n(\vec{r}, t) e^{-\I\omega t}
\end{eqnarray}

In order to obtain values of $\alpha$ at imaginary frequencies one
only has to substitute $\omega$ by ${\rm i}u$.  This computational
framework has been implemented in the {\tt octopus} code, described in
Ref.~\onlinecite{octopus-1}, to which we refer the reader
for technical details.  The {\tt octopus} code performs the
calculation on a real space mesh, which reduces the convergence
problem to two parameters: the grid spacing (we used 0.3~\AA\, in this
case) and the size of the simulation box.  The ion-electron
interaction was modelled with norm-conserving Troullier-Martins
pseudopotentials.\cite{troullier-1991}

The geometries of the PAHs were obtained by means of DFT as described
in Ref.~\onlinecite{malloci-2007-II}. For the optimization of the
geometries, we chose the hybrid B3LYP\cite{becke-1993} as the
approximation to the exchange and correlation functional. For the
TDDFT propagations, however, we used the adiabatic local density
approximation (LDA), which has proved to yield reliable results for
conjugated molecules;\cite{yabana-1999} the use of more sophisticated
functionals is possible, but does not change the results for the
dynamic polarizability significantly.\cite{marques-2001}

\section{RESULTS}

We have computed the $C^{AB}_6$ Hamaker constants for all possible
pairs $\lbrace A,B\rbrace$ of a set of 41 PAHs. The homo-molecular
Hamaker constants $C^{AA}_6$ for all the PAHs studied are reported in
Tables~\ref{table:pahsc6} and~\ref{table:pahsc6-2}; the full set can
be consulted in the database described in
Ref.~\onlinecite{malloci-2007-II}. The tables also display the average
static dipole polarizability $\alpha(0)$, the effective London
frequency $\omega_1$ (see below) and the retarded van der Walls
coefficient $K$.

It is very difficult to extract Hamaker constants experimentally; to
our knowledge, there are no experimental results reported for any PAH
to compare to. For benzene, however, Kumar and Meith\cite{kumar-1992}
reported a value of 1723\,a.u., by making use of the dipole oscillator
strength distributions (DOSDs), which are constructed from
experimental dipole oscillator strengths and molar refractivity
data. This is in good agreement with our computed value of 1763\,a.u.

We have also displayed for comparison the numbers reported in
Ref.~\onlinecite{jiemchooroj-2005}, obtained by means of the complex
linear polarization propagator scheme. This scheme was constructed on
top of TDDFT, although making use of the B3LYP functional. These
methodological differences, and further numerical details can explain
the very small differences, in all cases below 2\% -- see the results
in the Table~\ref{table:pahsc6} for benzene, naphthalene, anthracene
and tetracene (also known as naphthacene).

\begin{figure}

\centerline{\includegraphics[width=0.8\columnwidth]{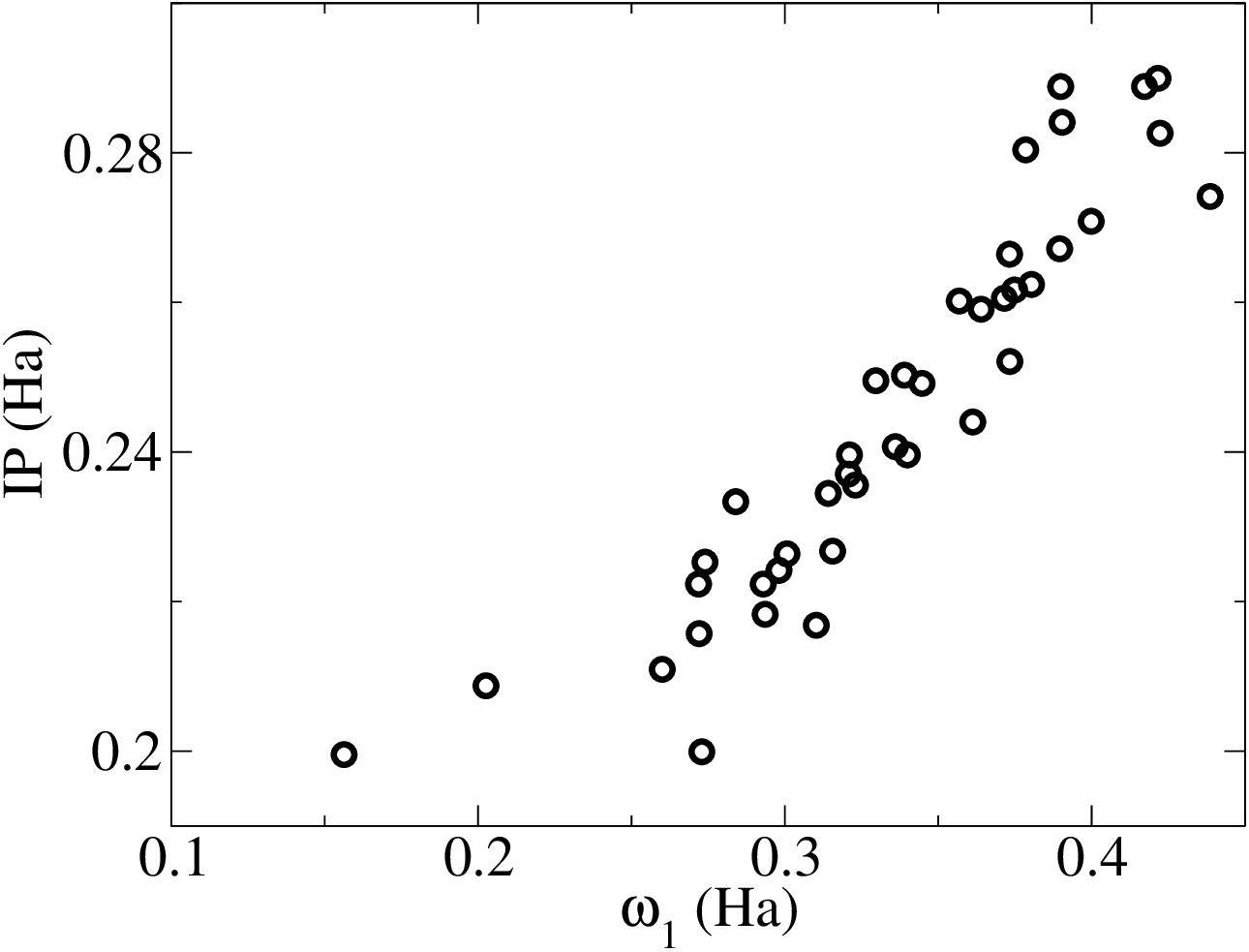}}

\vspace{24pt}

\caption{
\label{fig:w1}
Ionization potentials (vertical axis) against dispersion frequencies
(horizontal axis) for all the PAHs under study
}
\end{figure}

In the so-called London approximation, the polarizability at imaginary
frequencies is modelled with the help of only two parameters: the
static polarizability, and one effective frequency $\omega_1$:
\begin{equation}
  \alpha({\rm i}u) = \frac{\alpha(0)}{1+(u/\omega_1)^2}\,.
\end{equation}
Upon substitution on the Casimir-Polder integral, this yields for a
homo-molecular Hamaker constant:
\begin{equation}
\label{eq:w1}
  C_6 = \frac{3\omega_1}{4}\alpha^2(0)\,.
\end{equation}
With the knowledge of $C_6$ and $\alpha(0)$, one can obtain $\omega_1$
from Eq.~\ref{eq:w1}.  These effective frequencies are also reported
in Tables~\ref{table:pahsc6} and \ref{table:pahsc6-2}. They are
roughly decreasing with the size of the PAH, from 0.482~Ha for benzene
to the 0.156~Ha of pentarylene. The decrease is, however, strongly
irregular.  As it has been pointed out before\cite{mahan-1990}, it can
be related to the ionization potential of the molecules; this is
demonstrated in Fig.~\ref{fig:w1}, where we have plotted the
ionization potentials of the PAHs (calculated at the DFT/B3LYP level
of theory), versus the effective frequency $\omega_1$. The data points
approximately accumulate around a straight line, proving the
correlation.

\begin{figure}

\centerline{\includegraphics[width=0.9\columnwidth]{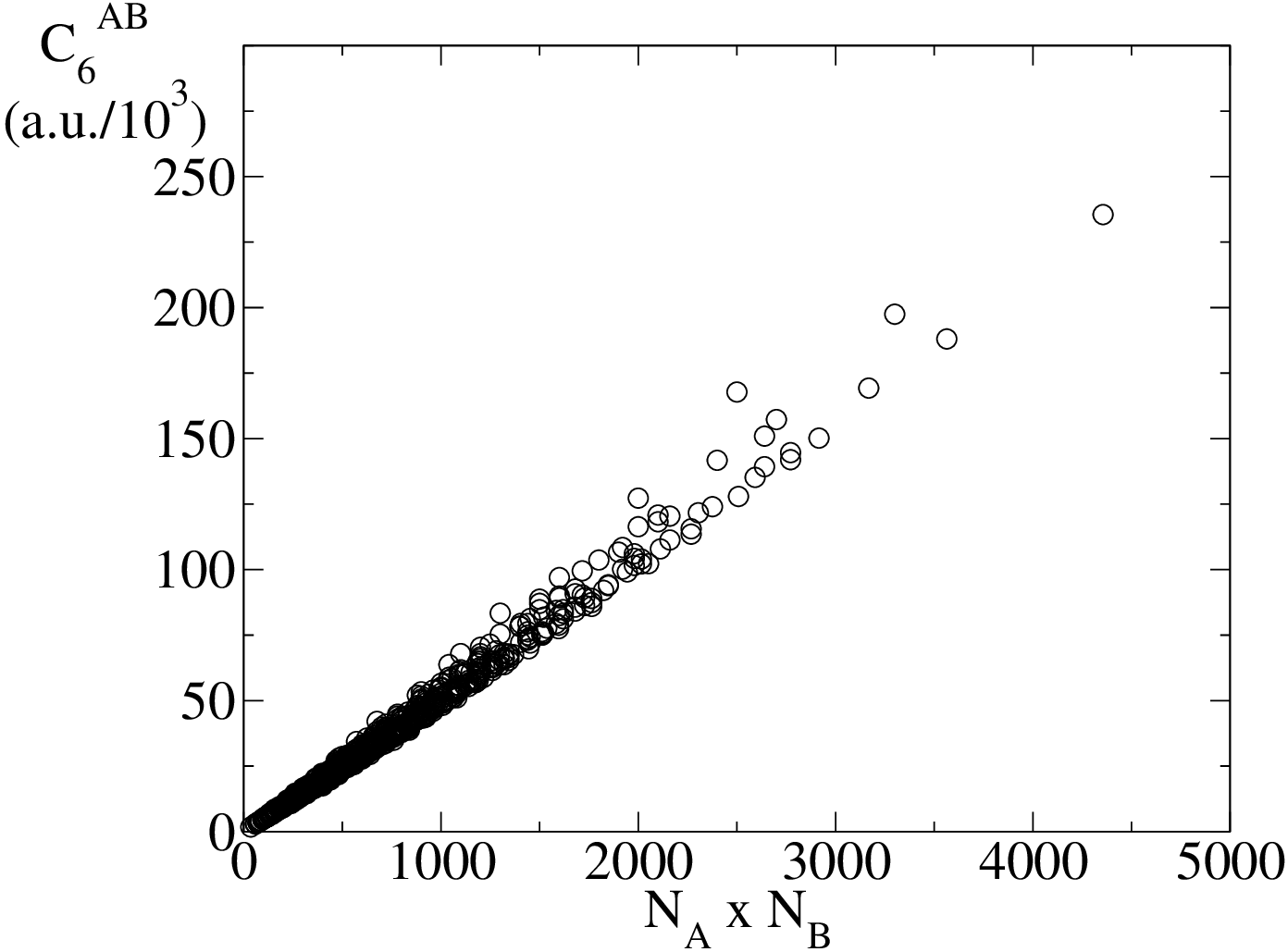}}

\vspace{24pt}

\caption{
\label{fig:pahs-natoms2}
$C_6^{AB}$ Hamaker constants for all the pairs of PAHs under study, as
a function of the product of the number of Carbon atoms, $N_A\times N_B$.  
}
\end{figure}

\begin{figure}

\centerline{\includegraphics[width=0.8\columnwidth]{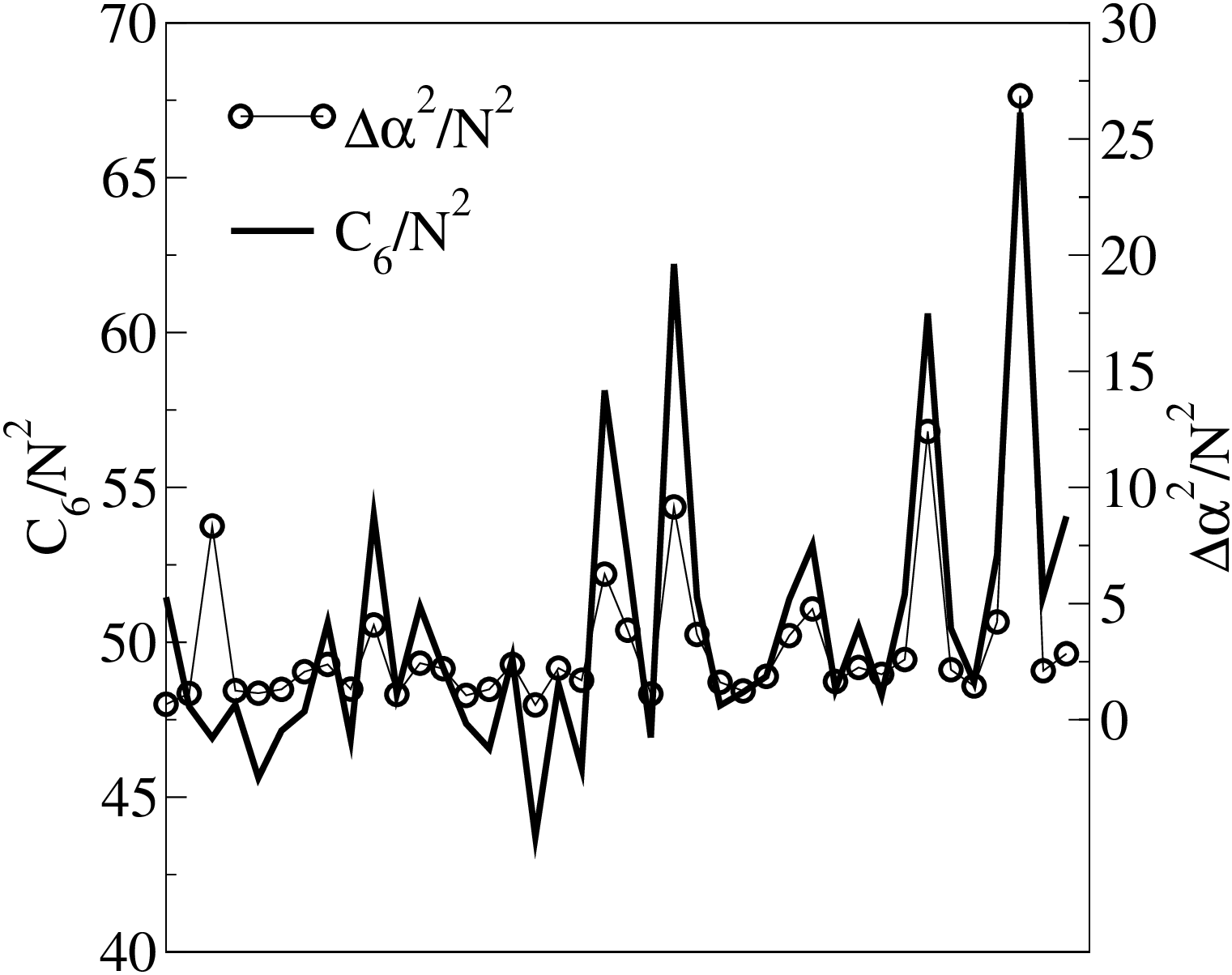}}

\vspace{24pt}

\caption{
\label{fig:pahs-deltaalpha}
$C_6$ homo-molecular Hamaker constants divided by the square of the
number of atoms (thick solid line, left axis), and dipole
polarizability anisotropy, also divided by the square of the number of
atoms (circles, right axis). Each data point corresponds with one PAH,
ordered in the $x$-axis according to its number of atoms (it is the
same ordering used in Tables~\ref{table:pahsc6}
and~\ref{table:pahsc6-2}).
}
\end{figure}

Equation~\ref{eq:w1} also gives us a hint on the dependency of $C_6$
with the size of the molecule: it is proportional to the square of the
polarizability (the product of polarizabilities, if the molecules are
different), which in turn typically grows with the volume, and
therefore, with the number of atoms. Consequently, one should expect a
linear dependency of $C_6^{AB}$ with respect to the product of the
number of atoms, $N_A \times N_B$. This is indeed confirmed in
Fig.~\ref{fig:pahs-natoms2}. Some cases, however, deviate from a
straight line. These cases correspond to strongly anisotropic PAHs
(with three very different axes). This is captured by the dipole
anisotropy:
\begin{equation}
  \Delta \alpha^2 = \frac{1}{3}[(\alpha_{xx}-\alpha_{yy})^2 +(\alpha_{xx}-\alpha_{zz})^2 + 
  (\alpha_{yy}-\alpha_{zz})^2]\,.
\end{equation}
Figure~\ref{fig:pahs-deltaalpha} shows how the PAHs whose Hamaker
constant deviates strongly from the general trend are those whose
polarizability anisotropy is also stronger: we have overlaid the
values of $C_6$ for all PAHs (divided by the square of the number of
Carbon atoms $N^2$), with the values $\Delta \alpha^2$ (also divided
by $N^2$). One can see how the two datasets are correlated --
specially in the right side of the graph, which corresponds to the
larger PAHs.

The crossover between the non-retarded and retarded regimes is given
by the length scale $\lambda = 2\pi c / \bar \omega$, where $\bar
\omega$ is a characteristic frequency of the electronic spectrum of
the molecule. For $R/\lambda \agt 10$ we enter the fully retarded
regime, while for $R/\lambda \alt 0.1$ we can still use
Eq.~\eqref{eq:eofr}. Using for $\bar \omega$ the values of $\omega_1$
obtained through the London approximation, we reach values in the
range of $\lambda \sim 0.1\,\mu$m (for benzene) to $\lambda \sim
0.3\,\mu$m (for pentarylene).

It is interesting to notice that in the fully retarded regime we can
write $K^{AA}$ as a function solely of the Hamaker $C^{AA}_6$
coefficient and the effective London frequency $\omega_1$. Combining
equations~\eqref{eq:kab} and \eqref{eq:w1} we arrive at
\begin{equation}
  K^{AA} = \frac{23 c}{6 \pi^2} \frac{C_6}{\omega_1} \,.
\end{equation}
The values of homomolecular coefficients $K^{AA}$ for the PAHs studied
in this work can be found in Tables~\ref{table:pahsc6}
and~\ref{table:pahsc6-2}.

\section{CONCLUSIONS}

In this Article we presented a method to calculate the van der Waals
coefficients of molecular systems using a time-propagation scheme
within TDDFT. As an example, we have applied this method to the family
of polycyclic aromatic hydrocarbons. Our results are in excellent
agreement with available theoretical and experimental data, and fully
validate our approach. Values of $C_6$ scale approximately with
$N_A\times N_B$, where $N_X$ is the number of atoms in the molecule
$X$. The strongest deviations from this law are for highly anisotropic
structures.

This scheme has several non-trivial advantages: (i)~It scales with
$N_a^2$, where $N_a$ is the number of atoms in the molecule. (ii)~The
time-propagation yields the polarizability in {\it real time}. From
this quantity it is then immediate to obtain the real and imaginary
parts of $\alpha$ at real and imaginary frequency. Therefore, the
optical absorption spectrum, static polarizabilities, $C_6$
coefficients, etc. are calculated in one shot. (iii)~This scheme is
easily generalized to higher order coefficients and arbitrary
geometries. We therefore expect it to be useful in the study of large
systems, like clusters or molecules with biological interest.

\section*{Acknowledgements}

We would like to thank A. Rubio for many useful discussions. M. A. L. Marques, A. Castro
and S. Botti acknowledge partial support by the EC Network of Excellence NANOQUANTA
(NMP4-CT-2004-500198). Part of the calculations were performed using
CINECA supercomputing facilities. G. Malloci acknowledges financial support
by RAS. G. Malloci and G. Mulas further acknowledge financial support by MIUR under
project PON-CyberSar. A. Castro acknowledges financial support from the
Deutsche Forschungsgemeinschaft within the SFB 450.

\clearpage

\begin{table}[t]
\caption{
\label{table:pahsc6}
Average static polarizability $\alpha(0)$, $C_6$ Hamaker constant,
effective frequency $\omega_1 = (4/3)C_6/\alpha^{2}(0)$, and the
coefficient $K$ of the retarded van der Waals interaction for the
selected PAHs.  All quantities are given in atomic units.  }
\begin{center}
\begin{tabular}{l|cccc}
Molecule                            & $\alpha(0)$ & $C_6$    & $\omega_1$ & $K$ \\
                                    &             & $\times 10^{-3}$ &    & $\times 10^{-6}$ \\
\hline
benzene                  (C$_{6}$H$_{6}$)     &   70.5    &  1.76  &  0.473  & 0.198     \\
{\footnotesize \hspace{12pt}(Ref.~\onlinecite{jiemchooroj-2005}; TDDFT/B3LYP)} &
       {\footnotesize 70.0}      & {\footnotesize 1.77}   & \footnotesize{0.482}  & \footnotesize{0.195}    \\
{\footnotesize \hspace{12pt}(Ref.~\onlinecite{kumar-1992}; DOSD)}              &
       {\footnotesize 67.8}      & {\footnotesize 1.72}   &     - & -             \\
azulene                  (C$_{10}$H$_{8}$)    &  133      & 5.15   & 0.390   & 0.703  \\
naphthalene              (C$_{10}$H$_{8}$)    &  123      & 4.79   & 0.422   & 0.605  \\
{\footnotesize \hspace{12pt}(Ref.~\onlinecite{jiemchooroj-2005}; TDDFT/B3LYP)} &
       {\footnotesize 122}      & {\footnotesize 4.87}   & \footnotesize{0.439}   & \footnotesize{0.594}    \\
acenaphthene             (C$_{12}$H$_{10}$)    &  143      & 6.76   & 0.439   & 0.820  \\
biphenylene              (C$_{12}$H$_{8}$)    &  152      & 6.91   & 0.400   & 0.920  \\
acenaphthylene           (C$_{12}$H$_{8}$)   &  145      & 6.57   & 0.417   & 0.838  \\
fluorene     margin            (C$_{13}$H$_{10}$)   &  159      & 7.97   & 0.422   & 1.00   \\
phenanthrene             (C$_{14}$H$_{10}$)   &  182      & 9.36   & 0.379   & 1.32   \\
anthracene               (C$_{14}$H$_{10}$)   &  189      & 9.92   & 0.372   & 1.42   \\
{\footnotesize \hspace{12pt}(Ref.~\onlinecite{jiemchooroj-2005}; TDDFT/B3LYP)} &
            {\footnotesize 185}      & {\footnotesize 10.0}   & \footnotesize{0.392} & \footnotesize{1.37}    \\
pyrene                   (C$_{16}$H$_{10}$)   &  205      & 12.0  & 0.380    & 1.68  \\
tetracene                (C$_{18}$H$_{12}$)   &  264      & 17.5  & 0.336    & 2.78  \\
{\footnotesize \hspace{12pt}(Ref.~\onlinecite{jiemchooroj-2005}; TDDFT/B3LYP)} &
            {\footnotesize 259}      & {\footnotesize 17.5}   & \footnotesize{0.349} & \footnotesize{2.68}     \\
triphenylene             (C$_{18}$H$_{12}$)   &  231      & 15.7  & 0.390    & 2.14  \\
benzo[a]anthracene       (C$_{18}$H$_{12}$)   &  246      & 16.6  & 0.364    & 2.42  \\
chrysene                 (C$_{18}$H$_{12}$)   &  239      & 15.9  & 0.373    & 2.27  \\
benzo[e]pyrene           (C$_{20}$H$_{12}$)   &  260      & 18.9  & 0.375    & 2.69  \\
perylene                 (C$_{20}$H$_{12}$)   &  262      & 18.6  & 0.361    & 2.74  \\
benzo[a]pyrene           (C$_{20}$H$_{12}$)   &  277      & 19.8  & 0.345    & 3.06  \\
corannulene              (C$_{20}$H$_{12}$)   &  244      & 17.5  & 0.390    & 2.38  \\
\end{tabular}
\end{center}
\end{table}

\clearpage

\begin{table}[t]
\caption{
  \label{table:pahsc6-2}
  Continuation of Table~\ref{table:pahsc6}.}
\begin{center}
\begin{tabular}{l|cccc}
Molecule                            & $\alpha(0)$ & $C_6$    & $\omega_1$  & $K$  \\
                                    &             & $\times 10^{-3}$ &    & $\times 10^{-6}$ \\
\hline
anthanthrene             (C$_{22}$H$_{12}$)   &  304      & 23.5  & 0.340  & 3.69  \\
benzo[g,h,i]perylene     (C$_{22}$H$_{12}$)   &  282      & 22.2  & 0.373  & 3.17  \\
pentacene                (C$_{22}$H$_{14}$)   &  353      & 28.1  & 0.301  & 4.98  \\
dibenzo[b,def]chrysene   (C$_{24}$H$_{14}$)   &  355      & 30.4  & 0.321  & 5.04  \\
coronene                 (C$_{24}$H$_{12}$)   &  318      & 27.0  & 0.357  & 4.03  \\
hexacene                 (C$_{26}$H$_{16}$)   &  454      & 42.1  & 0.272  & 8.23  \\
dibenzo[cd,lm]perylene   (C$_{26}$H$_{14}$)   &  384      & 34.8  & 0.314  & 5.89  \\
bisanthene               (C$_{28}$H$_{14}$)   &  402      & 37.6  & 0.310  & 6.45  \\
benzo[a]coronene         (C$_{28}$H$_{14}$)   &  386      & 37.9  & 0.339  & 5.96  \\
dibenzo[bc,ef]coronene   (C$_{30}$H$_{14}$)   &  431      & 44.0  & 0.316  & 7.42  \\
dibenzo[bc,kl]coronene   (C$_{30}$H$_{14}$)   &  459      & 46.3  & 0.293  & 8.40  \\
terrylene                (C$_{30}$H$_{16}$)   &  484      & 47.8  & 0.272  & 9.36  \\
ovalene                  (C$_{32}$H$_{14}$)   &  453      & 49.6  & 0.323  & 8.18  \\
tetrabenzocoronene       (C$_{36}$H$_{16}$)   &  565      & 65.4  & 0.273  & 12.8  \\
circumbiphenyl           (C$_{38}$H$_{16}$)   &  538      & 69.8  & 0.321  & 11.6  \\
circumanthracene         (C$_{40}$H$_{16}$)   &  612      & 82.5  & 0.293  & 15.0  \\
quaterrylene             (C$_{40}$H$_{20}$)   &  799      & 97.0  & 0.203  & 25.5  \\
circumpyrene             (C$_{42}$H$_{16}$)   &  631      & 89.0  & 0.298  & 15.9  \\
hexabenzocoronene        (C$_{42}$H$_{18}$)   &  590      & 85.9  & 0.330  & 13.9  \\
dicoronylene             (C$_{48}$H$_{20}$)   &  770      & 122 & 0.274  & 23.6  \\
pentarylene              (C$_{50}$H$_{24}$)   &  1196     & 168 & 0.156  & 57.1  \\
circumcoronene           (C$_{54}$H$_{18}$)   &  840      & 150 & 0.284  & 28.1  \\
circumovalene            (C$_{66}$H$_{20}$)   &  1099     & 236 & 0.260  & 48.2  \\
\end{tabular}
\end{center}
\end{table}

\clearpage

\listoffigures


\end{document}